\documentclass[journal]{IEEEtran}
 \usepackage{graphics,subfigure}

\usepackage{epsfig,xcolor}
\usepackage{array}
\usepackage{mdwmath}
\usepackage{mdwtab}
\usepackage{eqparbox}
\usepackage{multirow}
\usepackage[cmex10]{amsmath}
\usepackage{url}
\usepackage{graphicx}
\usepackage{subfigure}
\usepackage[absolute,showboxes]{textpos}
\usepackage{amssymb}
\begin{document}
\title{Hardware Implementation of Multimodal Biometric using Fingerprint and Iris}
\author{Tariq M. Khan   
\thanks{Tariq M. Khan is with Department of Engineering, Macquarie University, Sydney, Australia }}

\maketitle
\thispagestyle{empty}
\begin{abstract}
In this paper, a hardware architecture of a multimodal biometric system is presented that massively exploits the inherent parallelism. The proposed system is based on multiple biometric fusion that uses two biometric traits, fingerprint and iris. Each biometric trait is first optimised at the software level, by addressing some of the issues that directly affect the FAR and FRR. Then the hardware architectures for both biometric traits are presented, followed by a final multimodal hardware architecture. To the best of the author's knowledge, no other FPGA-based design exits that used these two traits.
\end{abstract}

\section{\bf \textsc{Introduction}}
In recent years, biometric authentication is gaining popularity because of its reliability and accuracy over the possession-based (e.g ID card) and knowledge-based (e.g pass code) authentication methods \cite{khan2016fusion}. Biometric identifiers can not be forgotten, guessed, misplaced or easily copied. However, despite the advantages of biometric authentication, biometric traits are facing numerous problems. These include inter-class similarity, intra-class variation, spoofing attacks and universality of the trait. Apart from these, it also suffers from enrollment problem because of noisy data resulting from defective sensors \cite{khan2019boosting}. Environmental variations, signal distortion, background noise, and the change in biometrics features can cause inherent variations in the biometric measurements. Therefore, a single biometric trait may not be sufficiently robust.\\
\indent A multimodal biometric system is introduced to overcome these problems. It uses multiple sensors to acquire biometric traits. This allows: (i) multiple units of same biometrics (middle and indexed fingerprints), multiple sensors of same biometrics (Capacitive and Optical fingerprint sensor), (iii) multiple representation and matching of same biometric (texture based or minutiae-based fingerprint ), (iv) multiple samples of same biometrics (three templates of left indexed fingerprint), and (v) multiple biometric (face and iris or fingerprint and iris). Because of this, a multimodal biometric system is less affected by noise, it overcomes the non-universality problem, it provides storage security environment and it improves the matching accuracy. Due to these advantages, it had received a considerable amount of attention from researchers.\\
\indent Most of the existing multimodal biometric systems are computer based. The authentication is performed in an insecure environment that uses the central server for template storage. This can cause a critical information leakage issue.  Another disadvantage of a multimodal system is it requires a large amount of processing as compared to a unimodal biometric system. This makes multimodal system less suitable for real-time application. Although, in multimodal biometric, most of the operations are independent. Because of serial nature of most of the programming languages, especially the one used in computers, these can not be performed at the same time. The implementation of a multimodal biometric system on hardware can address these critical problems.\\
\indent In this paper, we present the hardware architecture of a multimodal biometric that comprises of multiple biometric (fingerprint and iris). Proposed architecture provides massive exploitation of the inherent parallelism. Rest of the paper is organised as follows: Related work is discussed in section II. In section III, proposed software-based design for fingerprint feature extraction is discussed. Proposed software-based design for iris feature extraction is discussed in section IV.  Section V details the matching and fusion. Hardware implementation of proposed multimodal biometric system is presented in section VI. Experimental results are discussed in Section VII. Finally, section VIII presents our concluding remarks.
\begin{figure*}[h!]
  \centering
  \includegraphics[scale=0.8]{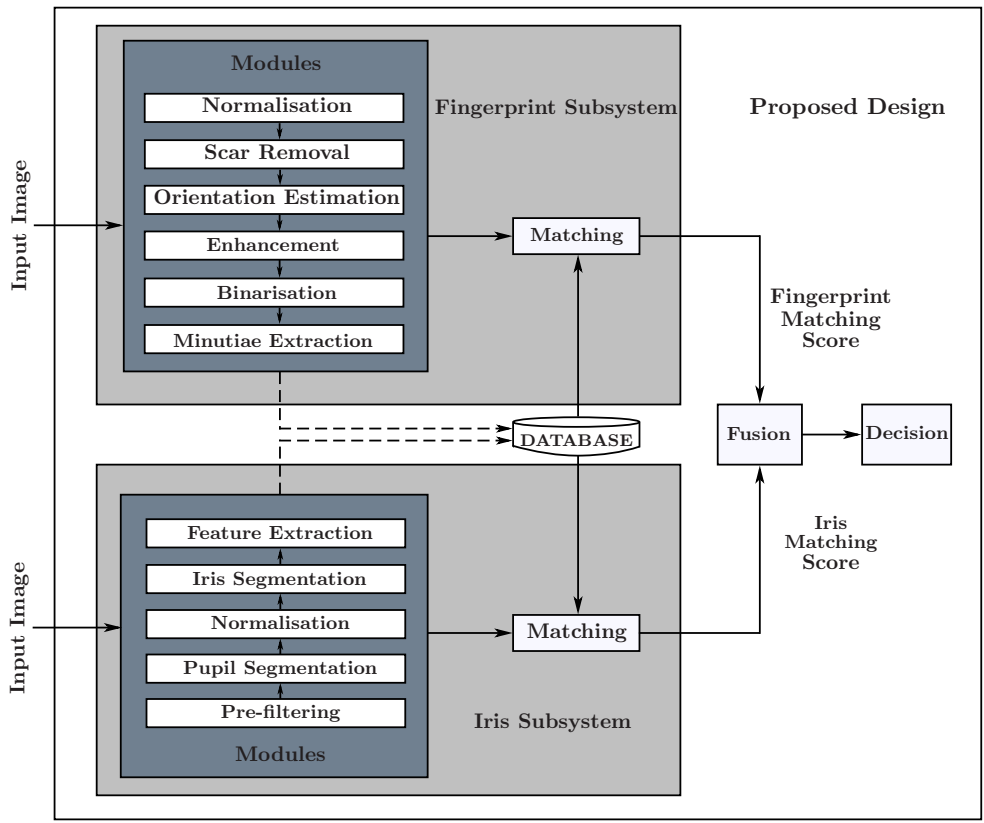}\\
  \caption{Block diagram of fingerprint feature extraction process }\label{FIS_Block}
\end{figure*}

\section{\bf \textsc{Related Work}}
From the literature, it is found that only a few multi-modal biometric systems are implemented as embedded systems. One reason is that a real-time embedded system in a resource-constrained environment poses great challenges, as it possesses limited computational resources and limited memory space. On the other hand, most of the existing multimodal biometric systems are computationally very expensive and are not suitable for real-time implementation. Converting the software design to hardware is one of the most difficult tasks. Therefore it is least developed, more so with fingerprint and iris multimodal biometrics. \\
\indent Sonal et al. \cite{Athalea2015} implement a palm-vein identification system in hardware. For hardware implementation a Blackfin ADSP-561 processor is used, whereas the C language is used for the algorithms used for matching of palm veins. Template matching and principal component analysis (PCA) are used as verification algorithms for palm-veins and are integrated at match score level. Yoo et al. \cite{Yoo2007} have developed two DSP systems for face-fingerprint and iris-fingerprint recognition. In their system, the most computationally expensive tasks are implemented on an FGPA in order to increase the system speed. They used a Xilinx XC3S4000 onboard FPGA and an ARM920T DSP clocked at 400 MHz, and a 128 MB SDRAM. However, no fusion strategy was applied in the embedded biometric system.\\
\indent Audrey et al. \cite{Poinsot2011} propose a contactless multimodal biometric system that combines two modalities: face and palmprint, by using fusion at the score level. This hardware architecture has been implemented on DSP and FPGA. Wang. J et al. ~\cite{Wang_09} proposed a multimodal biometric system that implements fingerprint and voiceprint. Matching-score level fusion was applied to voiceprint and fingerprint. They used an ARM9-Core based S3C2440A microprocessor that works at 400 MHz and the Microsoft Windows CE operating system. R. Moganeshwaran et al. ~\cite{Moganeshwaran_12} use finger vein and fingerprint for their multimodal biometric system. Two biometric traits, finger vein and fingerprint, are used and the whole process is implemented in SOC FPGA. The biometric fusion strategy applies at the matching score level. Conti et al. \cite{Contia2009} propose a multimodal technique for an embedded fingerprint recognizer. In this technique, fingerprint minutiae points along with fingerprint singularity points are used for robust user authentication. For biometric fusion, a matching score fusion module is used.

\section{Fingerprint Feature Extraction}
In fingerprint recognition system, reliable extraction of the minutiae (the ridge bifurcations and terminations) from the input fingerprint image is the most critical step that directly affects the recognition rate \cite{khan2010fingerprint,khan2013fingerprint,khan2014fingerprint,khan2016stopping}. The performance of minutiae extraction algorithm heavily depends on the quality of input image. This necessitates the use of good-quality input scans of fingerprints \cite{Khan2016,sabir2020reducing}. In reality, the acquired fingerprint images cannot be considered as good-quality scans in all circumstances. The presence of dirt or oil on the surface of the finger may result in a blurred scan. Furthermore, due to the electronic noise present in scanner electronics, the fingerprint scans may not be of clear quality. Mostly, the noise in a fingerprint image manifests itself in cuts or interrupted ridge lines. For an automatic fingerprint identification system (AFIS) to work reliably, these broken ridge lines need to be restored via an enhancement process \cite{Khan2016aaa}.\\
The minutiae extraction algorithm processes the fingerprint image in several stages in order to find the singular points related to bifurcation and termination of ridges. The number of
stages and the processing involved in each one differ slightly depending on the algorithm employed, being in our case five stages that are briefly described in this section.
\begin{figure*}[h!]
  \centering
  \includegraphics[scale=1]{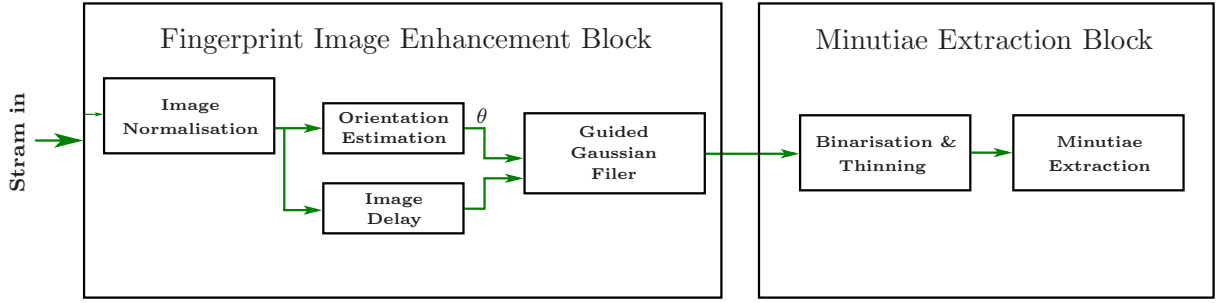}\\
  \caption{Block diagram of fingerprint feature extraction process }\label{FIS_Block}
\end{figure*}
\subsection{Image Normalization}
Normalisation is a process that changes the range of pixel intensity values. Normalisation is sometimes called contrast stretching or histogram stretching. In more general fields of data processing, such as digital signal processing, it is referred to as dynamic range expansion \cite{Gonzalez2006}. \cite{Lee1980}, and \cite{Mark2010} suggested local image statistics, such as mean and variance in a small neighbourhood, to incorporated in the contrast improvement strategy. The local normalisation method comprises of first dividing the image into appropriate small neighbourhoods and then normalising these neighbourhoods with respect to their local mean and variance. This will result in shaping these neighbourhoods to have a ridge/valley pattern with better contrast. Mathematically, it can be represented as
 \begin{equation}\label{NormEq}
 g\left( {x,y} \right) = \frac{{I\left( {x,y} \right) - {m_f}\left( {x,f} \right)}}{{{\sigma _f}\left( {x,y} \right)}}
 \end{equation}
where $I(x,y)$ is the input image, $m_f(x,y)$ is an estimation of a local mean of $I(x,y)$ and ${{{\sigma _f}\left( {x,y} \right)}}$ is an estimation of the local contrast (such as the standard deviation). Although the contrast is restored with no black patch at the centre, the amplitude of the granular noise in the background is significantly lifted. This happens due to the fact that the background area has almost zero local variance \cite{Singh2012}, thus resulting in the division a small number, which amplifies the noise structure. We propose a function of local variance which is used as a multiplying factor for the outcome of the second phase image. The factor is defined as
\begin{equation}\label{EQNORM111}
M=1 - \exp \left( { - \frac{{{\sigma _f^2}}}{{{2C^2}}}} \right)
\end{equation}
where $\sigma ^2_f$ is the local variance and $C$ is a user-defined parameter to regulate the noise suppression power in background areas. The value of $C$ is in the range 0-1, however, in our experiments, the value of 0.3 was adequate in all cases.

\subsection{Orientation estimation}
In this paper, the procedure outlined in \cite{Wang2007} was adopted for this purpose. First, discrete derivatives $G_x$ and $G_y$ in $x$ and $y$ directions are calculated by employing a Gaussian smoothed kernel, with a small standard deviation to mitigate noise. Then, covariance matrix data for the fingerprint image was calculated for each pixel as $G_{xx}= G_x^2$, $G_{xy}= G_x \times G_y$, and $G_{yy}= G_y^2$. The covariance matrix elements were further smoothed with a Gaussian having $\sigma=1$ standard deviation. Since a ridge line has two edges, the gradient vectors at both sides of a ridge are opposite to each other. If we want to calculate $\theta$ by taking the average of gradient angles directly in a local block, the opposite gradients at both sides of a ridge line are likely to cancel each other. To solve this problem, Kass and Witkin \cite{Kaas_87} proposed a simple and clever idea of doubling the gradient angles before the averaging process. These doubled angles are smoothed with a Gaussian of $\sigma=7$. Finally, the orientation is estimated by
\begin{equation}\label{Theta}
\theta  = \frac{\pi }{2} + \frac{{\arctan \left( {\frac{{cos\left( {2\theta } \right)}}{{sin\left( {2\theta } \right)}}} \right)}}{2}
\end{equation}
\subsection{Filtering}
\indent
In this paper, an oriented Gaussian filter is proposed by \cite{Khan2016a} is used that works similar to an anisotropic diffusion filter. However, for a general Gaussian filter, the separable axis do not align with the image axes. While separability can be used, it is more complex to implement in this case. It is even worse for a steerable filter, where the major axis of the Gaussian kernel changes with each pixel. The oriented Gaussian filter can be expressed as:
\begin{equation}\label{Eq2}
{G_{dir}}\left( {x,y;\theta ,f,{\sigma _x},{\sigma _y}} \right) = \exp \left\{ { - \frac{1}{2}\left( {\frac{{x_\theta ^2}}{{\sigma _x^2}} + \frac{{y_\theta ^2}}{{\sigma _y^2}}} \right)} \right\}
\end{equation}
To make the implementation process further simpler, $G_{dir}$ is decomposed into two filters. Since $\sigma _y^2 <  < \sigma _x^2$ when filtering ridge patterns (such as fingerprints), the filter can be decomposed into a small isotropic filter
\begin{equation}\label{Eq3}
{G_{iso}}\left( {y;{\sigma _y}} \right) = \exp \left\{ { - \frac{1}{2}\left( {\frac{{x_{}^2 + {y^2}}}{{\sigma _y^2}}} \right)} \right\}
\end{equation}
and an anisotropic 1D filter
\begin{equation}\label{Eq4}
{G_{ani}}\left( {{x_\theta };\theta ,{\sigma _\theta }} \right) = \exp \left\{ { - \frac{1}{2}\left( {\frac{{x_\theta ^2}}{{\sigma _\theta ^2}}} \right)} \right\}
\end{equation}
where $\sigma _\theta ^2 = \sigma _x^2 - \sigma _y^2 \approx \sigma _x^2$ and ${x_\theta } = x\cos \theta  + y\sin \theta $.
\subsection{Binarisation \& Thinning }
In binarisation, the grey scale image is converted to a binary image where the value of each pixel could be 1 or 0. Pixel set to 1 corresponds with a background/valley, whereas pixel set to 0 associated with foreground/ridge. As claimed by \cite{Khan2016b}, the normalisation facilitates the binarisation, therefore, a simple threshold works equally to the famous Otsu \cite{Otsu1979} thresholding method. The use of simple threshold makes the hardware implementation much easier than the existing thresholding methods. \\
\indent After binarisation, the next step is thinning is performed prior to minutiae extraction. For thinning, algorithm Zhang and Suen's modified by \cite{Sudiro2005} is used. In this process, 8 adjacent neighbours are evaluated to a central pixel that determines whether to delete this pixel or not. The original algorithm is modified and the process is based on the representation of the image with '1' for light (white) and '0' for dark (black) or region point is for pixel value '0' and background point is '1'.

\subsection{Minutiae Extraction}
After image pre-processing step, minutiae extraction process is applied. Minutiae point detection depends on pixel value ('0' or '1'). This process is quite simple as it can be carried out by examining the connectivity of the thinned image. Obtaining the parameters like type and position is quite easy \cite{Sudiro2012}. It depends on the position of pixel P its connectivity. If the connectivity is 1 then it corresponds to the End Point and if the connectivity is 3, it corresponds to Bifurcation Point (BP) of minutiae. To reduce false minutiae detected at the edge of the fingerprint image, the process to check candidate minutiae point whether at the edge of the image or not are applied. Checking the existence of pixel value '0' at the right, left, top or bottom of candidate minutiae points for specific distance does this process.
\section{Iris Feature Extraction}
The process of iris recognition can be mainly divided into three main subtasks:
\begin{itemize}
  \item The task of extracting the iris from an already acquired image i.e. Segmentation.
  \item The task of straightening of the extracted iris i.e. Normalization.
  \item The task of extracting the feature from normalized image i.e. Iris code making.
\end{itemize}
\begin{figure*}
  \centering
  \includegraphics[scale=1.2]{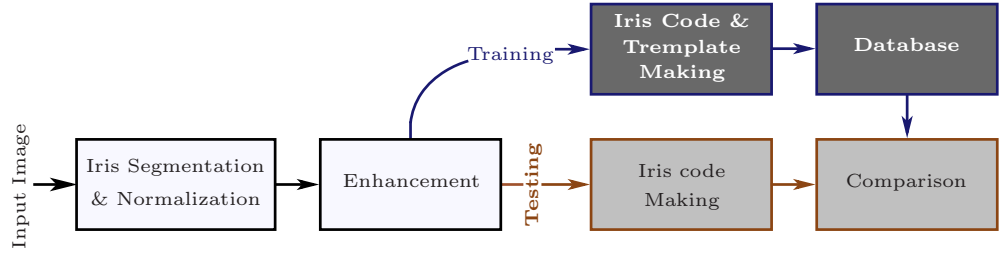}\\
  \caption{Block diagram of iris recognition system }\label{IrisBlock_New}
\end{figure*}
Fig.~\ref{IrisBlock_New} shows the block diagram of the proposed iris recognition system.
\subsection{Iris Segmentation and Normalisation}
n the field of Image Processing or specifically in computer vision, segmentation is defined as the partitioning of an image into smaller parts or segments which are easier to process and analyse \cite{Stockman2001,khan2011automatic, ibrahim2012iris}. In iris recognition, segmentation mean extraction the pupillary and the limbic boundary of an iris \cite{ibrahim2011novel}. This can be done either localising these boundaries simultaneously (e.g. Hough transform) or first locating the pupillary and then the limbic boundary (mostly using region property based methods). Hough transform based technique are iterative and are not suitable for hardware implementation whereas region properties based techniques are faster and suits for parallel processing. In region properties based techniques, the pupillary boundary is first located. Then using radial scanning (gradients) and interpolation the limbic boundary is located. Finally, the iris is normalised by using the information of segmented iris that requires interpolation again. Such a double conversion is challenging in a streamed hardware implementation. To overcome this problem Tariq et al. \cite{Khan2016c} uses the interpolation only a single time. In this paper, for iris segmentation and normalisation, we adapted same technique. In this technique, the pupillary boundary of the iris is first localised using a fast region property based method. Then, instead of locating limbic boundary, a region of interest is defined and the resultant image is converted from the rectangular to polar coordinates about the centre of pupil using

\begin{equation}\label{Eq1}
 X = r  \cos \left( {\frac{{\pi}}{{180}}} \right)
\end{equation}
\begin{equation}\label{Eq2}
Y = r  \sin \left( {\frac{{\pi}}{{180}}} \right)
\end{equation}
where r represents radius of the pupil.

After normalization, a gradient based method is used to locate the true limbic boundary.
\subsection{Feature Extraction/Iris code making }

\subsubsection{Image Enhancement}
\begin{figure*}
  \centering
  \includegraphics[scale=0.8]{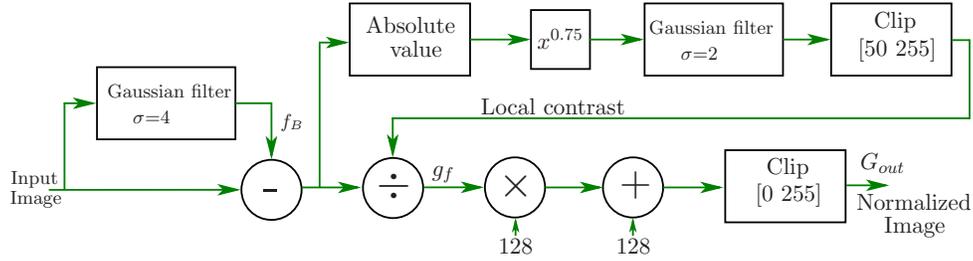}\\
  \caption{Image enhancement using local normalization}\label{LN}
\end{figure*}

\begin{figure}[h]
 	\begin{center}
 		\subfigure[] {\label{IrisEn:-a}\includegraphics[scale=0.37]{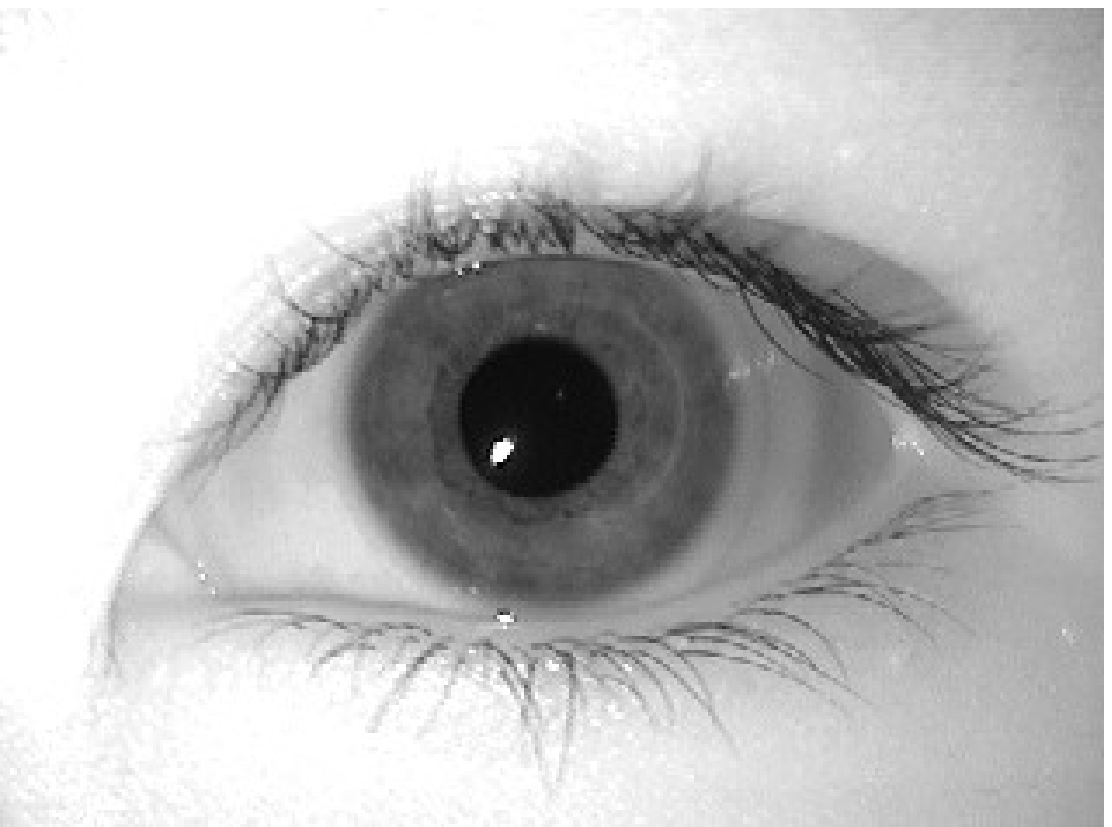}}
 		\subfigure[] {\label{IrisEn:-b}\includegraphics[scale=0.37]{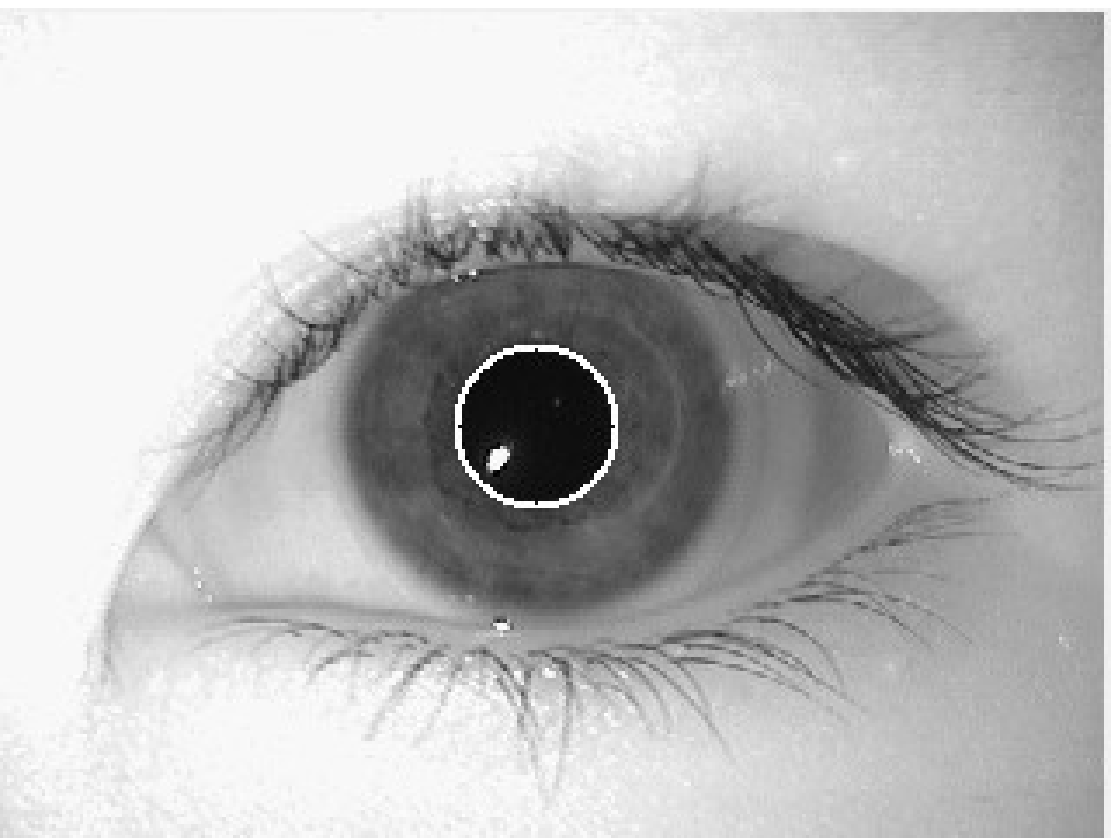}}
 		\subfigure[] {\label{IrisEn:-c}\includegraphics[scale=0.6]{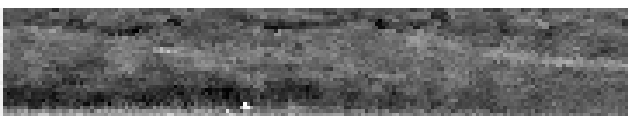}}
 		\subfigure[] {\label{IrisEn:-d}\includegraphics[scale=0.6]{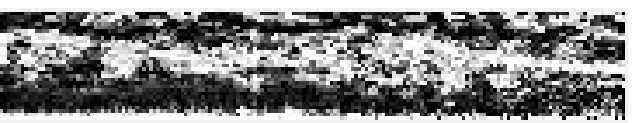}}
 	\end{center}
 	\caption{Iris segmentation and enhancement  (a) Sample eye image  (b) Pupil segmentation (c) Iris normalization and segmentation (d)  Iris feature enhancement. }
 	\label{IrisEn}
 \end{figure}

In iris code making, the normalised image is for enhanced by a contract normalisation process similar to the one proposed by \cite{Khan2016b}. This local normalisation technique deals with local image statistics in a better way. Another reason for selecting this technique is it best suits for hardware implementation. Fig.\ref{LN} shows the block diagram of the modified local normalisation technique. In this two phases are used; one that removes the non-uniform background and the other that restore the local contrast. In the first phase, the input image is subtracted from the Gaussian-weighted average smoothed by a low-pass Gaussian filter with $\sigma_1$. The parameter $\sigma_1$ can be set by utilising the fact that the filtered image should contain only the background changes (low-frequency content). In this paper, $\sigma_1=4$ is used. In the second phase, the local variance of the image is computed as an estimate of the local contrast. To normalise the contrast, the resultant image of the first phase is divided pixel-wise by the standard deviation of its spatial neighbours. Again, the size of the local variance filter depends on the size of the texture elements. As squaring the pixel value will double the base frequency, therefore, $\sigma_2$ is commonly taken to be $\sigma_1/2$. Fig.~\ref{IrisEn:-d} shows the enhanced iris image by proposed method.

\subsubsection{Bitplane Slicing}
In bitplane slicing, we first consider each grey level value in the image, in its binary equivalent form and then consider one of the eight bits at a time. For example, when considering bit plane 0, we check the least significant bit of each value while forcing all other bits to zero. Now, if the least significant bit is 1, we replace the whole number by 1 grey level value in the image and if it is zero, we replace the whole number by zero grey level value. In this way, bit plane zero is formed for the input image.\\
\indent Similarly, when considering bit plane 1, we consider the second bit from the right side of the binary sequence while forcing all others to zero. In this case, however, if the bit under consideration is 1, the original grey level value is replaced by 2 because so we get the binary representation in 0 and 1 form. This process of thresholding continues in this manner for all the seven bit planes.\\
 \begin{figure}[h]
 	\begin{center}
 		\subfigure[] {\label{pupil:-a}\includegraphics[scale=0.7]{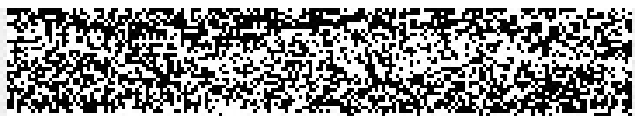}}
 		\subfigure[] {\label{pupil:-b}\includegraphics[scale=0.7]{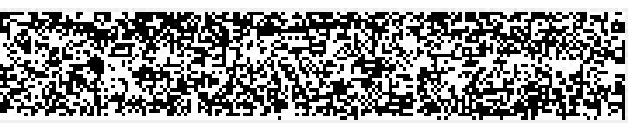}}
 		\subfigure[] {\label{pupil:-c}\includegraphics[scale=0.7]{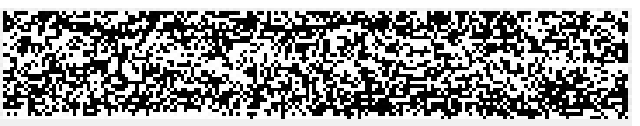}}
 		\subfigure[] {\label{pupil:-d}\includegraphics[scale=0.7]{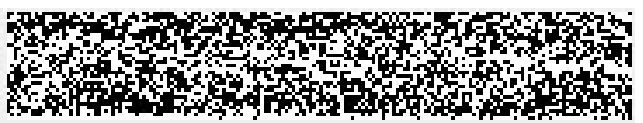}}
  		\subfigure[] {\label{pupil:-a}\includegraphics[scale=0.7]{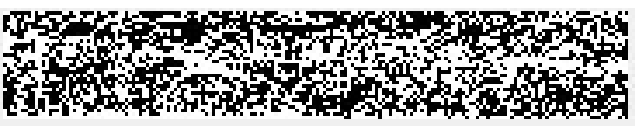}}
 		\subfigure[] {\label{pupil:-b}\includegraphics[scale=0.7]{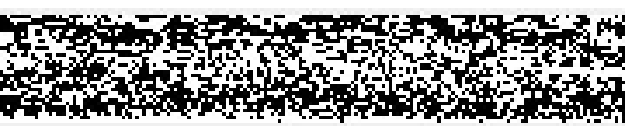}}
 		\subfigure[] {\label{pupil:-c}\includegraphics[scale=0.7]{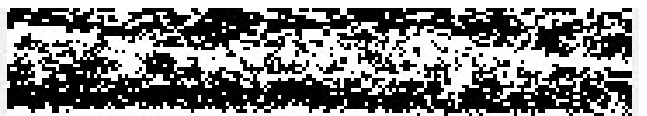}}
 		\subfigure[] {\label{pupil:-d}\includegraphics[scale=0.7]{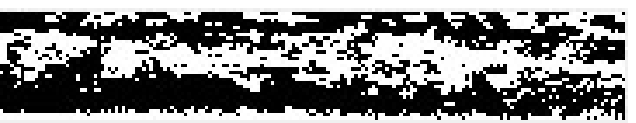}}
 	\end{center}
 	\caption{Bit Plane slicing: (a) Bit 0.  (b) Bit 1. (c) Bit 2. (d)  Bit 3 (e) Bit 4.  (f) Bit 5 . (g) Bit 6. (h)  Bit 7. }
 	\label{BitPlanSlice}
 \end{figure}
By applying bitplane slicing on the 8-bits of the image grey level values, we get 8 slices, as shown in Fig.~\ref{BitPlanSlice}. For iris code generation in our project, we have used the concept of bitplane slicing \cite{Basit2007}. We generate 8 slices of our enhanced normalised image as mentioned above.\\
\indent In this paper, six of the eight generated slices or planes are used to implement the bitplane slicing on the enhanced normalised image. Bit planes 0 and 7 are discarded since it is clear from the histogram of the iris image that the lowest and highest grey level component do not fall in the iris region of the eye, mostly the middle grey level values comprise it. The remaining bitplanes are termed as a feature of type 1, 2, 3 and so on. This method of representing the features in binary codes makes the comparison process more efficient.
\subsubsection{Iris code making}
As mentioned above, we have used 6 bitplanes or slices and have neglected the bit planes 0and 7. By considering the fact that each bit plane has two values either 0 or a non-zero value, we have normalised all the non-zero values to 1 in all bitplanes, thus generating binary codes, so that a matching code can easily be generated. Then by using the 6 bitplanes normalised to 0's and 1's, we generate a code for size Mx6, where M is the length of 1D transformed row vector of the normalised iris image. The final matrix is generated by simply merging the codes of the 5 bitplanes vertically. For testing hardware, 5 people of MMU-v1 iris database is used. This database has 5 left and 5 right eyes for each person and the database comprises a total of 44 persons. For iris code generation, three left and three right eye images are used while the remaining 2 left and 2 right eye images are used for testing purposes. The iris code is generated according to the following criteria.\\

1.	Three images are used for generating each person database
2.	Check the majority bit, which is selected as either 0 or 1 by comparing the code. For example, by comparing the codes of a person's left eye1, left eye2 and left eye3, each of size M×5 generated by the method mentioned above.

\begin{figure*}
  \centering
  \includegraphics[scale=1]{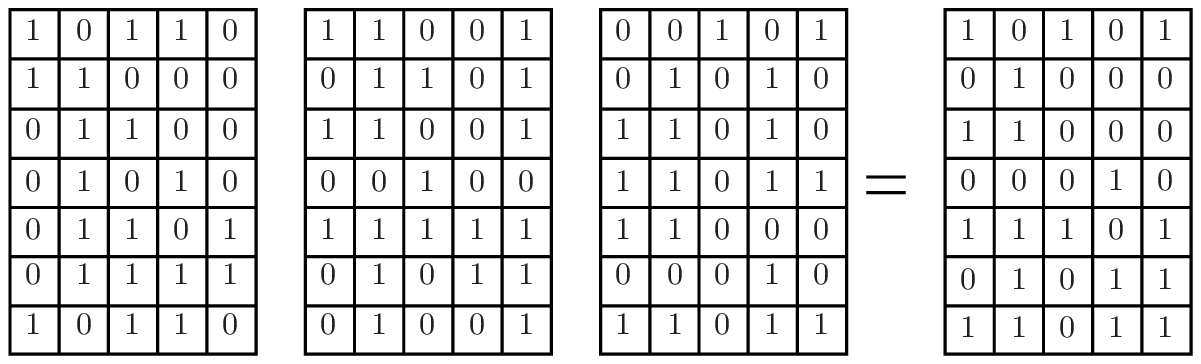}\\
  \caption{Iris code generated by applying majority bit selection}\label{MBS}
\end{figure*}
The whole process is well illustrated in Fig.\ref{MBS} where we have 3 left eye images in matrix form which are normalised to 0 and 1 after applying the bitplane slicing. The fourth matrix  the result of the method mentioned above and this matrix is later on used for comparison after storing it in the database.
\section{Matching and Fusion}
The proposed approach is based on a pair recognition of fingerprint and iris, and every part provides its own Matching score.  In fingerprints, minutiae matching is based on two stages: fingerprint alignment and fingerprint matching. In fingerprint alignment, a pair of minutiae ( one form input minutiae and one from the template minutiae) are located and their polar location in the polar coordinates is located that is relative to the pair of aligned minutiae. For this purpose, we followed the technique proposed by Lindoso et. al. \cite{Lindoso2007}. The output of this step give each minutiae a triplet representation: $(r,\theta,o)$ , where $\theta$ is radial angle, $r$ is radial distance and $o$ is relative orientation. Finally, all the minutiae data set of the input image are compared with all the minutiae date set of the template. To compensate the errors of location the comparison uses an adaptive elastic algorithm. The final match ratio for two fingerprints is stored.  \\
\indent After creating the biometric vectors The homogenous biometric vector from fingerprint and iris data is composed of binary sequences representing the unimodal biometric templates.  Matching score for Iris is calculated through Hamming distance (HD) between two final fused templates.
\begin{equation}\label{Eqfusion}
 HD = \frac{1}{N}\sum\limits_{i - 1}^N {XOR\left( {T{r_{ic}}\left( i \right),T{s_{ic}}\left( i \right)} \right)}
\end{equation}
where $T{r_{ic}}$ is the training feature vector and $T{s_{ic}}$ is the feature vector of test image.\\

Once the matching scores of both biometric traits are obtained then these are fused using simple sum rule. If the Fused matching score is larger than a pre-specified threshold, then the parson is accepted or rejected.
\section{\bf \textsc{Hardware Implementation}}
In software, usually one operation is performed at a time and its result is stored in RAM for the next operation. This is the reason why it takes a longer time to perform a certain task which comprises of multiple sequential operations. While on the hardware, these components can be combined to create parallel computing structures \cite{Bailey2011}. Almost all image processing algorithms contain operations that execute in sequence. This is a form of temporal parallelism \cite{Bailey2011}. Hence, this structure is ideal to have a separate processor for each operation. This is also known as a pipelined architecture. When processing images, data can usually begin to be output from an operation long before the complete image has been processed by that operation. The time between when data is first input to an operation and the corresponding output is available is the latency of that operation. When each operation only uses input pixel values from a small, local neighbourhood then its latency is lowest. This is because each output only requires data from a few input pixel values. Operation pipelining can give significant performance improvements when all of the operations have low latency because a downstream processor may begin performing its operation before the upstream processors have completed.
\subsection{Fingerprint Feature Extraction and Matching}
\begin{figure*}[htbp]
 	\begin{center}
 			\includegraphics[width=6in]{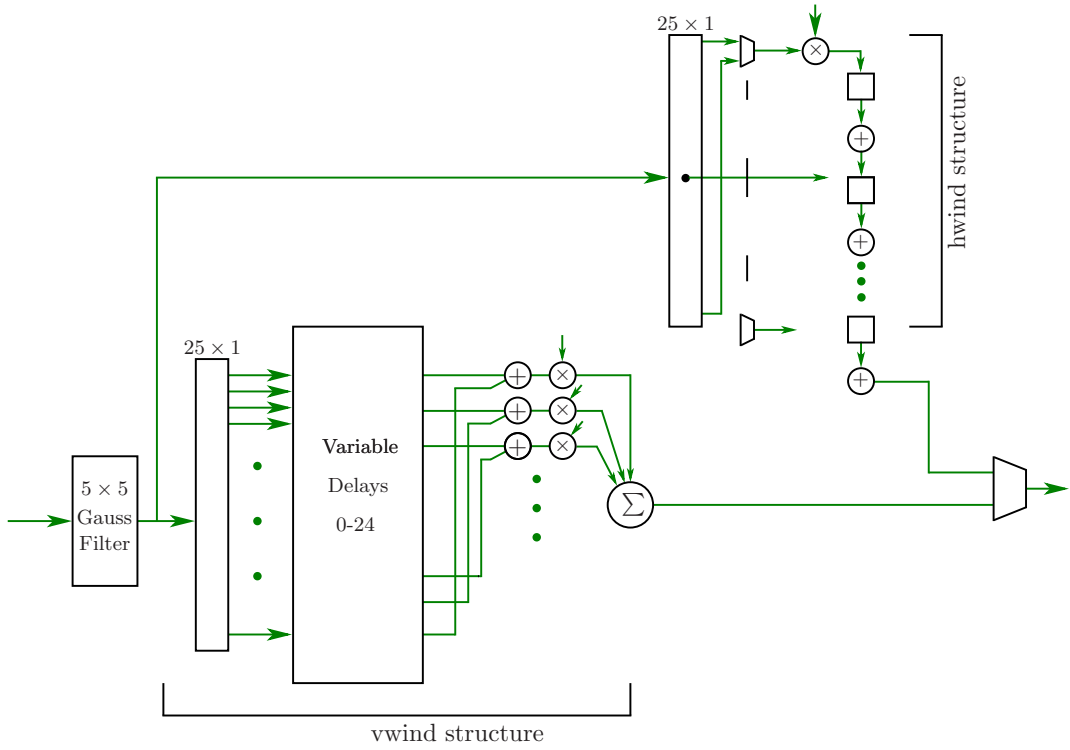}
 	\end{center}
 	\caption{Proposed guided line Gaussian structure}
 	\label{TunableGaussFilter}
 \end{figure*}

\begin{figure*}[htbp]
 	\begin{center}
 			\includegraphics[width=6in]{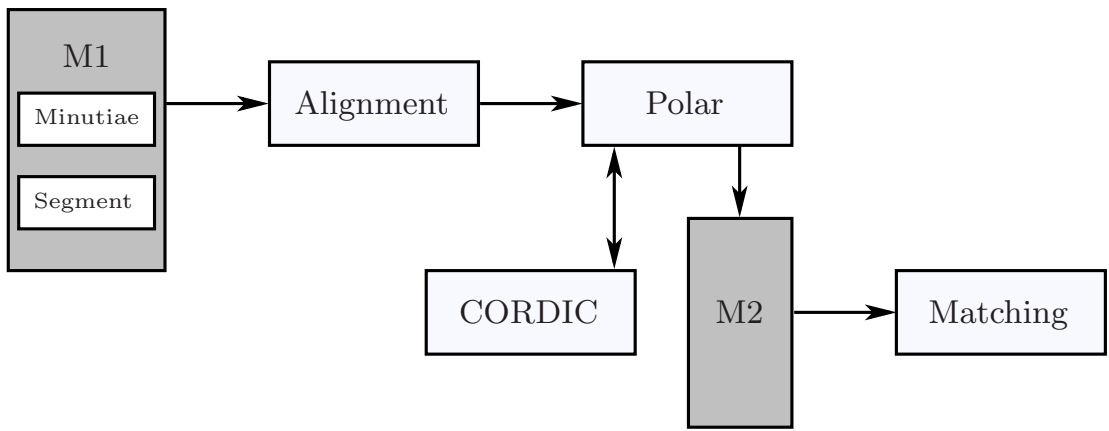}
 	\end{center}
 	\caption{Hardware structure for fingerprint alignment algorithm}
 	\label{MinuAlign}
 \end{figure*}

For fingerprint feature extraction first input image is enhanced. For image enhancement, a dynamically steerable Gaussian filter proposed by \cite{Khan2016a} is used. It is observed that with $\sigma_x=4$ the width of the line Gaussian can be reduced from 25 to 17 pixels ($\pm2\sigma$) without any significant. For this size window, this enables a simpler nearest neighbour interpolation to be used which significantly reduces the hardware complexity. To convert the 2-D filter into 1-D, the window is divided into two sub-windows hwind and vwind, which filter angles that are primarily horizontal and vertical respectively, as shown in Fig.~\ref{TunableGaussFilter}. Angles within vwind require one pixel from each row within the window, while those in hwind require one from each column. The pixels corresponding to the required delays are selected and then multiplied with the Gaussian weights. Finally, these are summed up to get the final resultant value. After enhancement, the fingerprint thinning process becomes quite easy.  In this process, 8 adjacent neighbours are evaluated to a central pixel that determines whether to delete this pixel or not. The minutiae feature bifurcation and ending are obtained by a cross numbering approach \cite{Sudiro2012}. In hardware, this approach is easy to implement with several gates like addition, subtraction, and shift registers.
For fingerprint alignment and matching, a pre-alignment algorithm is used \cite{Lindoso2007}. Fig.~\ref{MinuAlign} shows the hardware structure foe this pre-alignment algorithm. To implement this, two memories are required. First memory M1 is consists of two sub-memories to stores the extracted minutiae and related segments like position information and angle. While M2 store the minutiae of polar coordinates. In all memories, for angles and coordinates, 8 bites are required while for minutiae only 1 bit is required. In alignment block, best pair of minutiae is searched. All the input minutiae are aligned by using the difference of position and angle between two best pairs. The second block computes the modulus and angles using CORDIC. Finally, matching block compares all the aligned minutiae in polar coordinates.

\subsection{Iris Feature Extraction and Matching}
\begin{figure*}
  \centering
  \includegraphics[scale=1]{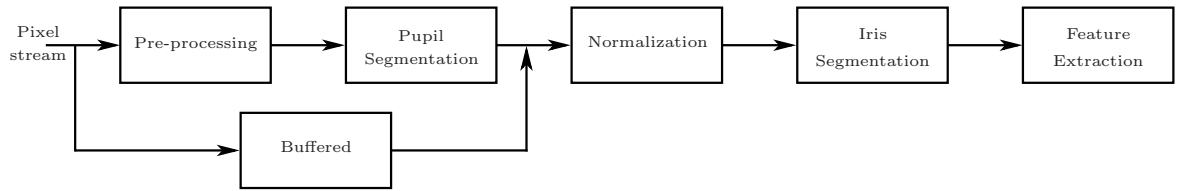}\\
  \caption{Block diagram of iris feature extraction process}\label{IFE_Block}
\end{figure*}

The proposed iris feature extraction is based on five steps, as shown in Fig.~\ref{IFE_Block}. The input image is first pre-processed by a mean subtraction. In mean subtraction operation, the image is smooth through a 2-D Gaussian filter of $\sigma=5$. The output of this Gaussian filter is subtracted from delayed image. The image is then thresholded at 0 by keeping only sign bit. Two morphological operators erosion and dilation are applied on this binary image. The resultant image is then scanned for connected component analysis. Two region properties area and eccentricity are used to isolate the pupil region from another region of the eye image. Once the centre and radius of the pupil are located the image is cropped that contain both pupil and iris region. This cropped image is buffered in SRAM. Using bilinear interpolation the image is normalised. Then using the first order vertical gradient operator the limbic boundary is isolated that give the normalised iris.\\
\indent After the normalisation, the image features are enhanced by using a local image normalisation. For local normalisation, the background is estimated by subtracting the mean image from the input image. For local contrast estimation, the magnitude is obtained by applying the absolute operator. Then the dynamic compression is done by using power-law transformation with $\gamma=0.75$ is applied to compress the high contrast. The resultant image is averaged locally with another Gaussian. Then by clipping the local contrast into the range [50-255] noise is suppressed. The resultant image is divided by the local contrast and the output is scaled to 128 and offset by 128.  After enhancement, the next step is to implement the bit plane slicing and create a feature vector. Both steps are quite easy to perform in hardware as we already deal in bits in hardware. Finally, the input feature vector is compared with the already stored vector. For testing purposed we only store 5 persons feature in SRAM.

\subsection{Fusion}
For fusion, the Matching score of two biometric traits sum rule is used \cite{Patil2013}. This requires normalizing the scores before combining them. The reason is, both biometric traits are of different nature. The normalisation transforms the score into a common range between 0 and 1\cite{Patil2013}. Finally, the score is summed and the decision is made on the base of the threshold.
\section{\bf \textsc{Experimental Results and Discussion}}
For fusion, the Matching score of two biometric traits sum rule is used \cite{Patil2013}. This requires normalising the scores before combining them. The reason is, both biometric traits are of different nature. The normalisation transforms the score into a common range between 0 and 1\cite{Patil2013}. Finally, the score is summed and the decision is made on the base of the threshold. \\

Two well-known parameters FAR and FRR are used for the performance evaluation. FAR is the number of times that the access occurred for an incorrectly accepted unauthorised person. FRR is a number of times that an incorrectly rejected authorised access. First, a test has been conducted on full FVC2002 DB2A database using our proposed minutiae-based recognition system. This resulted in FAR 1.27\% and FRR 18.38\%. For MMU v1 database \cite{mmu}, the proposed method resulted in FAR 4.29\% and FRR 15.77\%.\\

For the multimodal test, a database is created that consists 30 people of selected fingerprint database and 30 people of selected Iris database. Classical fusion technique matching-score level is used for fusion. Euclidean metric is applied to the HD of each subsystem. With the proposed approach for the multimodal biometric system, the following results have been obtained: FAR 1.97\% and FRR 12.79\% . Literature shows that fingerprint-based systems have fewer accuracies than iris-based systems \cite{Prabhakar1998, Ratha2000}. For this reason, we give higher weight to iris than fingering ( 0.4 for fingerprint and 0.6 for iris).\\

For hardware implemented a low-cost Cyclone IV GX P4CGX150F31FPGA is used. This FPGA combines an Intel embedded processor with Altera Cyclone IV GX FPGA. This is a full-featured computer system which is used for software hardware co-designs. In Table~\ref{MMBT1} the detail of hardware resource utilisation of the proposed iris recognition system is presented. It can be observed that the proposed method only consumes almost 12K Logic elements almost 12 \% of the total logic elements available. Our design also consumes about 10\% logic register. In Table~\ref{MMBT1} the detail of hardware resource utilisation of the proposed fingerprint recognition system is presented. If we compare the Logic element consumption of both biometric traits then proposed fingerprint recognition system consumes more logic cell and logic register than the iris recognition. On the other hand, the proposed iris recognition system consumes more memory bits than the fingerprint recognition system. The reason is, in iris recognition the whole image needs to be buffered once for extracting the outer boundary. Also, the template size of the iris is much bigger than the fingerprint that requires more memory to store fiord recognition.\\

In Table~\ref{MMBT3}, the relative processing speed of the MATLAB-based proposed fingerprint recognition system, iris recognition system and fusion is compared with its FPGA-based structure. The proposed fingerprint recognition is over 240 times faster than the MATLAB-based implementation on a PCA. The proposed iris recognition is over 197 times faster than the MATLAB-based implementation. Our complete multimodal biometric system takes about 16 seconds to recognise a person. In FPGA implemented, both biometric traits processed in parallel that significant boost the overall speed of the system. Our hardware-based multimodal system takes about 60 milliseconds to recognise a person that is over 270 times faster than the MATLAB-based system. The reason for this high speed is efficiently used of parallelism in the FPGA.

\begin{table*}[h!]
\centering
\caption{ Detailed Hardware resource utilization of the proposed iris recognition system on a low cost Cyclone IV GX FPGA}
\begin{tabular}{p{2cm}p{1.4cm}p{2.6cm}p{1.6cm}p{2.1cm}p{2cm}}\hline\\
Resources                                     & Available     &  Pupil Segmentation & Normalization   & Iris Segmentation and Enhancement & Iris recognition \\\hline
&&&&\\
Logic Elements                            & 149760            & 2810      & 597       & 2986      & 5251   \\
&&&&\\
Logic register                                & 149760            & 1174      & 335       & 1100      & 3212       \\
&&&&\\
Memory bits                                 & 6MB               & 612k     & 364k      & 88k         & 386k    \\
&&&&\\\hline
\end{tabular}
\label{MMBT1}
\end{table*}

\begin{table*}[h!]
\centering
\caption{ Detailed Hardware resource utilization of the proposed fingerprint recognition stem on a low cost Cyclone IV GX FPGA}
\begin{tabular}{p{2cm}p{1.4cm}p{2.6cm}p{1.6cm}p{1.6cm}p{1.6cm}}\hline\\
Resources                                     & Available     &  Normalization & Orientation estimation   &  Filtering & Minutiae recognition \\\hline
&&&&\\
Logic Elements                            & 149760            & 2286      &  7597       & 8268      & 4297   \\
&&&&\\
Logic register                              & 149760            & 917         & 3335       & 5334      & 2235       \\
&&&&\\
Memory bits                                 & 6MB               & 60k       & 164k         & 58k         & 86k    \\ &&&&\\\hline
\end{tabular}
\label{MMBT2}
\end{table*}
\begin{table}[h!]
\caption{ Processing speed of proposed FPGA based algorithm with proposed PC based MATLAB structure}
\centering
\begin{tabular}{p{3cm}p{2cm}p{1cm}p{1cm}} \hline\\
FVC2004 Database Type & Proposed MATLAB(PC) & Proposed FPGA& Speedup \\ \hline
&&&\\
Fingerprint recognition  &   8.95     &  0.0360 & 240$\times$  \\
&&&\\
Iris recognition  &   5       &  0.0260 & 197$\times$   \\
&&&\\
Fusion  &   2.24       &   0.0120&188$\times$  \\&&& \\  \hline
\end{tabular}
\label{MMBT3}
\end{table}
\vspace{-0.1cm}

\section{\bf \textsc{ Conclusion}}
This paper presents a reliable multimodal biometric system that respects multiple constraints: low-cost, real- time processing, hygienic, straightforwardness, user-friendliness, limited memory, etc. To achieve this, we present a hardware architecture of a multimodal biometric system that massively exploits the inherent parallelism. The proposed system is based on multiple biometric fusion that uses two biometric traits, fingerprint and iris. Both fingerprint and iris are highly accurate biometric traits. The proposed system id efficiently implemented in hardware. As far as the authors know, the proposed structure is the only one that gives the hardware implementation of a complete multimodal biometric using fingerprint and iris recognition. The proposed  hardware system is over 270 times faster than the MATLAB-based system.We also plan to investigate the further optimisation of the both biometric traits to improve the FAR and FRR and the further optimise the hardware resource utilizations.

\bibliographystyle{IEEEtran}
\bibliography{references}

\end{document}